\documentclass[letter,twocolumn]{jpsj3}
\usepackage[usenames]{color}

\def\blue#1{{\color{black}#1}}
\def\vec#1{\mib #1}
\def\Ham{\mathcal H}
\def\Hsq{\Ham^{\mathrm{sq}}}
\def\vS{\vec{S}}
\def\vtau{\vec{\tau}}
\def\vtone{\vec{\tau}^{(1)}}
\def\vttwo{\vec{\tau}^{(2)}}
\newcommand{\ket}[1]{\left\vert {#1} \right\rangle}
\def\DM{\ket{\mathrm{DM}}}
\def\Jm{J_{\mathrm{m}}}

\title
{
Dimer-Monomer Ground State for Extended Spin-1/2 Diamond Chain
}

\author
{Ken'ichi {Takano}\thanks{E-mail: takano@toyota-ti.ac.jp}
}

\inst
{
Toyota Technological Institute, 
Tenpaku-ku, Nagoya 468-8511, Japan}

\recdate
{June 19, 2017
}

\abst
{
We present a condition \blue{in which} the dimer-monomer state is exactly the ground state of an extended diamond chain with spin magnitude 1/2. 
The Hamiltonian \blue{of the extended diamond chain} includes next-nearest-neighbor exchange interactions and distortions, 
\blue{where} the spin magnitude of the spin pair on a singlet dimer \blue{is} not generally conserved. 
The method \blue{of} deriving the condition is based on representing the Hamiltonian in a complete square form. 
The dimer-monomer ground state is found even \blue{if} the Hamiltonian \blue{has no} space-reflection symmetries. 
}

\begin{document}

\sloppy

\maketitle

\begin{figure} %Fig.1
\vspace{10 mm}
\begin{center} 
\includegraphics[width=0.75 \linewidth]{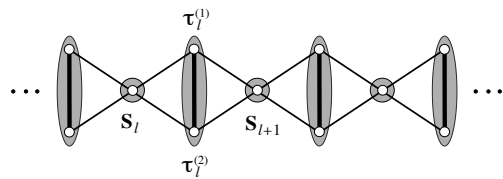}
\end{center}
%\vspace{-1 zw}
\caption{Original diamond chain and the DM state. 
An open circle represents a half spin and a bold or thin line represents an exchange interaction. 
\blue{The DM state consists of singlet dimers (shaded ovals) and free spins (shaded circles).}
}
\label{diamond_chain}
\end{figure}

Spin systems with frustration in low dimensions are interesting, since they produce various types of quantum spin liquids. 
The diamond chain depicted in Fig. \ref{diamond_chain} is a frustrated spin system which has several different spin-liquid ground states along with the change of a parameter. 
The full ground-state phase diagrams are found by both rigorous and numerical analyses for the $S$=1/2 and 1 diamond chains.\cite{tks,ht2017} 
In particular, the dimer-monomer (DM) state is commonly a ground state for \blue{the} diamond chains of any spin magnitudes $S$; 
we know the exact form,\cite{tks} which is shown with the lattice in Fig.~\ref{diamond_chain}. 
The DM state is characteristic in that it includes effective free spins, monomers, in strong exchange interactions. 

When a distortion is introduced to the Hamiltonian representing the $S$=1/2 diamond chain, detailed numerical analysis is performed.\cite{ottk,otk} 
Under distortion, the exact DM state is \blue{no longer} the ground state but is continuously changed to a gapless spin-liquid state without free spins. 
Experimentally, azurite Cu$_3$(CO$_3$)$_2$(OH)$_2$ is found to be a substance in such a spin-liquid state for a distorted diamond chain;\cite{kfcm,kfcm_sup} in particular, a 1/3 magnetic plateau is observed. 
Cu$_3$Cl$_6$(H$_2$O)$_2\cdot$2H$_8$C$_4$SO$_2$ \blue{is also found to represent} a distorted diamond chain, \blue{which has an excitation gap.\cite{ithu}
} 
Theoretical analyses including arguments on \blue{the} values of exchange parameters under distortion are given.\cite{hl} 
Recently, substances for extended versions of diamond chains \blue{were} found.\cite{fkmh,ftfk,mfks}
They include next-nearest-neighbor exchange interactions as well as distortions. 
The exchange interaction between monomers \blue{were} theoretically and experimentally considered.\cite{jokv,ots,fkmm} 
\blue{The} search for substances realizing extended diamond chains are going to develop. 
In view of the present situation, more information about extended diamond chains is expected to be brought. 

In this Letter, we investigate the extended $S$=1/2 diamond chain \blue{for which} the DM state is the ground state. 
The method we use is based on representing the Hamiltonian in a complete square form. 
This method \blue{was} developed from the method \blue{that uses} projection operators.\cite{t1994jpa,t1994jps}
In particular, Ref. \citen{t1994jpa} is the first paper to treat exactly the diamond chain in a special case, where the method of complete square form is applied at the phase boundary between the DM phase and the tetramer-dimer (TD) phase. 

\begin{figure} %Fig.2
\begin{center} 
\includegraphics[width=0.55 \linewidth]{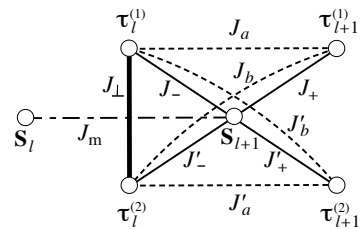}
\end{center}
%\vspace{-1 zw}
\caption{Exchange parameters in a unit cell. 
They include next-nearest-neighbor exchange interactions and distortions.
}
\label{unit_extend}
\vspace{-5 mm}
\end{figure}

\blue{The exchange} interactions \blue{that} we consider is shown in Fig.~\ref{unit_extend}. 
They \blue{are} all the interactions in a simple choice of unit cell. 
Then the Hamiltonian for this extended $S=1/2$ diamond chain is written as 

\begin{align}
\label{HamOrg}
\Ham = \sum_l 
&\Big(  J_{\perp} \vtone_{l} \!\cdot \vttwo_{l} 
\nonumber\\[-2.5 mm]
& + J_{-} \vtone_{l} \!\cdot \vS_{l+1} 
+ J'_{-} \vttwo_{l} \!\cdot \vS_{l+1} 
\nonumber\\
& + J_{+} \vtone_{l+1} \!\cdot \vS_{l+1} 
+ J'_{+} \vttwo_{l+1} \!\cdot \vS_{l+1} 
\nonumber\\
& + J_{a} \vtone_{l} \!\cdot \vtone_{l+1} 
+ J'_{a} \vttwo_{l} \!\cdot \vttwo_{l+1} 
\nonumber\\
& + J_{b} \vttwo_{l} \!\cdot \vtone_{l+1} 
+ J'_{b} \vtone_{l} \!\cdot \vttwo_{l+1} 
\nonumber\\
& + J_{\mathrm{m}} \vS_{l} \!\cdot \vS_{l+1} \Big) , 
\end{align}
where $\vtau^{(1)}_{l}$, $\vtau^{(2)}_{l}$, 
and $\vS_{l}$ are spin operators with magnitude 1/2 in the $l$th unit cell. 
The sum, with respect to $l$, is taken over $N$ unit cells with \blue{a} large $N$ limit, 
and \blue{boundary effects are not considered}. 
For this Hamiltonian, $\vec{T}_l^2$ with $\vec{T}_l \equiv \vtone_{l}+\vttwo_{l}$ is not generally conserved.

The DM state is written as 
\begin{align}
\DM = \bigotimes_l \ket{a_l} \otimes \ket{0_l} , 
\label{DMstate}
\end{align}
where $\ket{a_l}$ is any state of $\vS_{l}$ and $\ket{0_l}$ is the singlet state of $\vtone_{l}$ and $\vttwo_{l}$. 
There are infinite degenerate DM states owing to the arbitrariness of $\ket{a_l}$'s, and $\DM$ represents one of them.

\begin{figure} %Fig.3
\vspace{10 mm}
\begin{center} 
\includegraphics[width=0.6 \linewidth]{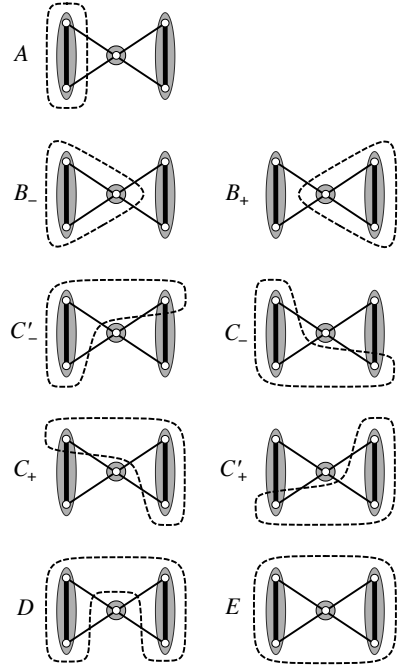}
\end{center}
\caption{All the possible \blue{groupings} of spins in the DM state. 
The grouping rule is that the spins in a group form a partial eigenstate with \blue{a} total spin magnitude \blue{of} 0 or 1/2 for the DM state, and 
any two of \blue{the} spins in a group are connected by an exchange interaction in Fig. \ref{unit_extend}.}
\label{sq_grouping}
%\vspace{-2 zw}
\end{figure}

To analyze $\Ham$, we construct another Hamiltonian $\Hsq$ in a complete square form. 
$\Hsq$ is the linear combination of the squares of the spin summations of some spins except for a constant term. 
It includes the terms of any spin \blue{group} satisfying the rule: 
the spins in a group form a partial eigenstate with \blue{a} total spin magnitude \blue{of} 0 or 1/2 for the DM state, and any two of spins in a group are connected by \blue{one of exchange interactions shown} in Fig. \ref{unit_extend}. 
All the possible grouping are represented in Fig.~\ref{sq_grouping}. 
Then the Hamiltonian is 
\begin{align}
\label{HamSq}
\Hsq &= \sum_l 
\bigg\{ \frac{1}{2} A \, \left(\vtone_{l}+\vttwo_{l}\right)^2 
\nonumber\\ 
&+ \frac{1}{2} B_{-} \left[ \left(\vtone_{l}+\vttwo_{l} + \vS_{l+1} \right)^2 - \frac{3}{4} \right] 
\nonumber\\ 
&+ \frac{1}{2} B_{+} \left[ \left(\vtone_{l+1}+\vttwo_{l+1} + \vS_{l+1} \right)^2 - \frac{3}{4} \right] 
\nonumber\\ 
&+ \frac{1}{2} C'_{-} \left[ \left(\vtone_{l}+\vttwo_{l} + \vtone_{l+1} \right)^2 - \frac{3}{4} \right] 
\nonumber\\ 
&+ \frac{1}{2} C_{-} \left[ \left(\vtone_{l}+\vttwo_{l} + \vttwo_{l+1} \right)^2 - \frac{3}{4} \right] 
\nonumber\\ 
&+ \frac{1}{2} C_{+} \left[ \left(\vtone_{l+1}+\vttwo_{l+1} + \vtone_{l} \right)^2 - \frac{3}{4} \right] 
\nonumber\\ 
&+ \frac{1}{2} C'_{+} \left[ \left(\vtone_{l+1}+\vttwo_{l+1} + \vttwo_{l} \right)^2 - \frac{3}{4} \right] 
\nonumber\\ 
&+ \frac{1}{2} D \left(\vtone_{l}+\vttwo_{l} + \vtone_{l+1}+\vttwo_{l+1} \right)^2 
\nonumber\\ 
&+ \frac{1}{2} E \left[ \left(\vtone_{l}+\vttwo_{l} + \vtone_{l+1}+\vttwo_{l+1} + \vS_{l+1} \right)^2 - \frac{3}{4} \right] \bigg\} 
\nonumber\\ 
&+ U_{0} 
\end{align}
with 
\begin{align}
U_{0} &= - \frac{3}{4} ( A + B_{-} + B_{+} + C'_{-} + C_{-} 
\nonumber\\ 
&\qquad\qquad + C_{+} + C'_{+} + 2D + 2E ) N , 
\label{GeneCoef}
\end{align}
where $A, B_{-}, B_{+}, C'_{-}, C_{-}, C_{+}, C'_{+}, D$, and $E$, are constant coefficients. 
Each term in $\Hsq$ vanishes except for $U_0$, if it operates on an eigenstate of the lowest spin magnitude, 0 or 1/2, for the total spin within the term. 
Hence the DM state (\ref{DMstate}) is the ground state of $\Hsq$, if all the coefficients, $A, B_{-}, B_{+}, C'_{-}, C_{-}, C_{+}, C'_{+}, D$, and $E$, are nonnegative. 
Then the ground-state energy is $U_{0}$, since all the terms other than $U_{0}$ \blue{have} the minimum value, zero. 

For $\Hsq$, there is no ground state \blue{other than} the DM state in the following cases: 
(a) $A$ is positive; 
(b) one of $B_{-}$ and $B_{+}$, and one of $C'_{-}, C_{-}, C_{+}$, and $C'_{+}$ are positive; and 
(c) two of $C'_{-}, C_{-}, C_{+}$, and $C'_{+}$ are positive. 
In case (a), $A>0$ enforces the two spins, $\vtone_{l}$ and $\vttwo_{l}$, to form a singlet dimer in the ground state. 
Then, all $\vS_{l}$s become free, since there is no interaction between $\vS_{l}$ and $\vS_{l+1}$ for all $l$. 
As for cases (b) and (c), we consider only $C'_{-}$ is positive, for example. 
Then, there are two ground states: 
one is the DM state, and the other is the state in which $\vttwo_{l}$ and $\vtone_{l+1}$ form a dimer for each $l$. 
Another positive coefficient, $B_{-}, B_{+}, C_{-}, C_{+}$, or $C'_{+}$, excludes the latter. 
\blue{When} only $B_{-}$ and $B_{+}$ are positive, \blue{the DM state is not the only ground states.} 
In fact, we can locally replace the dimer of $\vtone_{l}$ and $\vttwo_{l}$ and the monomers of $\vS_{l}$ and $\vS_{l+1}$ in the DM state with two dimers of $\vtone_{l}$ and $\vS_{l}$ and of $\vttwo_{l}$ and $\vS_{l+1}$ without  increasing the energy. 

The DM state is the ground state of $\Ham$, if $\Ham = \Hsq$ and all the coefficients in $\Hsq$ are nonnegative. 
\blue{When} $\Ham = \Hsq$, by expanding $\Hsq$ and comparing \blue{the} coefficients, we have 
\begin{subequations}
\label{CoefJ}
\begin{align}
A + B_{+} + B_{-} &+ C_{+} + C'_{+} \nonumber\\
+ \, C'_{-} &+ C_{-} + 2D + 2E = J_{\perp} , 
\label{CoefJperp}\\
B_{-} + E &= J_{-} = J'_{-} , 
\label{CoefJminus}\\
B_{+} + E &= J_{+} = J'_{+} ,  
\label{CoefJplus}\\
C'_{-} + C_{+} &+ D + E = J_{a} , 
\label{CoefJa}\\
C_{-} + C'_{+} &+ D + E = J'_{a} , 
\label{CoefJad}\\
C'_{-} + C'_{+} &+ D + E = J_{b} , 
\label{CoefJb}\\
C_{-} + C_{+} &+ D + E = J'_{b} , 
\label{CoefJbd}\\
0 &= \Jm . 
\label{CoefJm}
\end{align}
\end{subequations}
These equations mean that the DM state is an eigenstate of $\Ham$ owing to $\Ham=\Hsq$. 
In (\ref{CoefJ}), we directly \blue{observe the following} restrictions \blue{in} exchange parameters: 
\begin{subequations}
\label{JJdJm}
\begin{align}
J'_{+} &= J_{+} , 
\label{JpJdp}\\
J'_{-} &= J_{-} , 
\label{JmJdm}\\
\Jm &= 0 . 
\label{Jmzero}
\end{align}
\end{subequations}
\blue{Further,} the consistency of \blue{equations} (\ref{CoefJa}) to (\ref{CoefJbd}) provides another restriction 
\begin{align}
J_{a} + J'_{a} = J_{b} + J'_{b} . 
\label{JadaJbdb}
\end{align}
Then \blue{the equations} (\ref{CoefJa}) to (\ref{CoefJbd}) are not independent. 
Hence, we exclude (\ref{CoefJbd}) hereafter. 
We also know that the nonnegative coefficients require nonnegative values for the exchange parameters, owing to (\ref{CoefJ}). 
By using (\ref{CoefJperp}), the energy of the DM state (\ref{GeneCoef}) is written as 
\begin{align}
U_{0} = - \frac{3}{4} J_{\perp} N . 
\label{GeneJ}
\end{align}

We now have 6 linear simultaneous equations, (\ref{CoefJperp}) to (\ref{CoefJb}), for 9 variables, $A, B_{-}, B_{+}, C'_{-}, C_{-}, C_{+}, C'_{+}, D$, and $E$. 
We take $C_{+}$, $C_{-}$, and $E$ as arbitrary numbers. 
Then the solution is written as 
\begin{subequations}
\label{CoefEq}
\begin{align}
A &= J_{\perp} - J_{+} - J_{-} - J_{a} - J'_{a} + 2E , 
\label{CoefEq_a}\\ 
B_{+} &= J_{+} - E , 
\label{CoefEq_bp}\\ 
B_{-} &= J_{-} - E , 
\label{CoefEq_bm}\\ 
C'_{+} &= - J_{a} + J_{b} + C_{+} , 
\label{CoefEq_cdp}\\ 
C'_{-} &= - J'_{a} + J_{b} + C_{-} , 
\label{CoefEq_cdm}\\ 
D &= J_{a} + J'_{a} - J_{b} - E - C_{+} - C_{-} . 
\label{CoefEq_d}
\end{align}
\end{subequations}

To (\ref{CoefEq}), we apply the other condition that all \blue{the} coefficients are nonnegative. 
\blue{This results in the following} inequalities, 
\begin{subequations}
\label{Jcond}
\begin{align}
J_{\perp} - J_{+} - J_{-} &- J_{a} - J'_{a} \ge -2E , 
\label{Jcond_a}\\
J_{+} &\ge E , 
\label{Jcond_b}\\
J_{-} &\ge E , 
\label{Jcond_c}\\
J_{a} - J_{b} &\le C_{+} , 
\label{Jcond_d}\\
J'_{a} - J_{b} &\le C_{-} , 
\label{Jcond_e}\\
J_{a} + J'_{a} - J_{b} &\ge E + C_{+} + C_{-} , 
\label{Jcond_f}
\end{align}
\end{subequations}
where $C_{+}$, $C_{-}$, and $E$ are arbitrary nonnegative numbers. 

In conclusion, the Hamiltonian (\ref{HamOrg}) has the DM ground state (\ref{DMstate}), if 
(i) the exchange parameters are all nonnegative, 
(ii) (\ref{JJdJm}) and (\ref{JadaJbdb}) are satisfied, and 
(iii) there exist nonnegative numbers, $C_{+}$, $C_{-}$, and $E$, 
satisfying (\ref{Jcond}). 
Then the ground state energy is \blue{given by} (\ref{GeneJ}). 
Further, there is no ground state except for the DM state, if 
(a) $A$ is positive; 
(b) one of $B_{-}$ and $B_{+}$, and one of $C'_{-}, C_{-}, C_{+}$, and $C'_{+}$ are positive; or 
(c) two of $C'_{-}, C_{-}, C_{+}$, and $C'_{+}$ are positive. 
The positivities are known by (\ref{CoefEq}). 

We examine, as a simple example, the case that (\ref{JJdJm}) is satisfied and 
$J_{a} = J'_{a} = J_{b} = J'_{b} = 0$. 
\blue{Then,} (\ref{JadaJbdb}) is satisfied. 
(\ref{Jcond_f}) requires $C_{+} = C_{-} = E = 0$. 
Then \blue{equations} (\ref{Jcond_b}) to (\ref{Jcond_e}) are satisfied. 
(\ref{Jcond_a}) reduces to 
\begin{align}
J_{\perp} \ge J_{+} + J_{-} . 
\label{Jspecial}
\end{align}
\blue{The equations} (\ref{CoefEq_cdp}) and (\ref{CoefEq_cdm}) deduce $C'_{+} = C'_{-} = 0$ so that the conditions (b) and (c) are not satisfied. 
As to the condition (a), (\ref{CoefEq_a}) mentions that only the DM state is the ground state if $J_{\perp} > J_{+} + J_{-}$. 
In the symmetric case of $J_{+} = J_{-}$, (\ref{Jspecial}) reproduces the \blue{entire} range of the DM phase for the original spin-1/2 diamond chain in Ref. \citen{tks}. 
This type of distorted diamond chain is examined by the coupled cluster method.\cite{jlty} 

We also examine another example to see that the condition for the DM ground state has a nontrivial solution. 
Our example is the case that (\ref{JJdJm}) is satisfied and $J_{\perp} = 2.09$, $J_{+} = 0.9$, $J_{-} = 1.1$, $J_{a} = 0.05$, $J'_{a} = 0.05$, $J_{b} = 0.01$, and $J'_{b} = 0.09$ in \blue{an} arbitrary energy unit. 
First, (\ref{JadaJbdb}) is satisfied. 
Since (\ref{Jcond_a}) reduces to $E \ge 0.005$, we take $E = 0.005$ for trial. 
Then (\ref{Jcond_b}) and (\ref{Jcond_c}) are satisfied. 
Since (\ref{Jcond_d}) and (\ref{Jcond_e}) reduce to $C_{+} \ge 0.04$ and $C_{-} \ge 0.04$, respectively, we try to take $C_{+} = C_{-} = 0.04$. 
Then (\ref{Jcond_f}) becomes $0.09 \ge 0.085$, which is a consistent inequality. 
Therefore the DM state is the ground state. 
\blue{Further}, since $C_{+}$ and $C_{-}$ are positive, only the DM state is the ground state. 

\newpage

To summarize, we obtained a sufficient condition that the DM state is the ground state for the extended diamond chain (\ref{HamOrg}); 
the method deriving the condition is based on representing the Hamiltonian in a complete square form. 
The DM state is the ground state for a wide range of nonsymmetric Hamiltonians with next-nearest-neighbor exchange interactions and distortions. 
They do not generally have the conservation law of $(\vtone_{l}+\vttwo_{l})^2$ and the space-reflection symmetries with respect to the site of $\vS_l$ and to the line of $\vtone_{l}$ and $\vttwo_{l}$. 
Experimentally, substances satisfying (\ref{JJdJm}) and (\ref{JadaJbdb}) are not found in \blue{literature}. 
However, \blue{with} the recent discovery of diamond-chain substances, it \blue{can be hoped} that substances satisfying the present condition will be found in \blue{the} near future.

\acknowledgment
I would like to thank Kazuo Hida for discussions. 
This work is  supported by JSPS KAKENHI Grant Number JP26400411.

\end{document}